\documentclass[prd,twocolumn,twoside,preprintnumbers,nofootinbib,natbib,bm,amsrefs]{revtex4-2}

\usepackage{hyperref,url}
\usepackage{color,xcolor,graphicx,tabularx,tabularray}
\usepackage{natbib}
\usepackage[compat=1.1.0]{tikz-feynman}

\definecolor{lg}{RGB}{220,220,220} 
\definecolor{lime}{HTML}{A6CE39}

\DeclareRobustCommand{\orcidicon}{\hspace{-2.1mm}
\begin{tikzpicture}
\draw[lime,fill=lime] (0,0.0) circle [radius=0.13] node[white] {{\fontfamily{qag}\selectfont \tiny ID}}; \draw[white,fill=white] (-0.0525,0.095) circle [radius=0.007]; 
\end{tikzpicture} \hspace{-3.7mm} }
\foreach \x in {A, ..., Z} {\expandafter\xdef\csname orcid\x\endcsname{\noexpand\href{https://orcid.org/\csname orcidauthor\x\endcsname} {\noexpand\orcidicon}}}

\newcommand{\Eprint}[1]{\href{#1}}

\begin{document}

\preprint{PSI-PR-23-4, ZU-TH 09/23, ICPP-70}

\title{Searching for low-mass resonances decaying into $W$ bosons}

\author{Guglielmo Coloretti\orcidA{}}
\email{guglielmo.coloretti@physik.uzh.ch}
\affiliation{Physik-Institut, Universität Zürich, Winterthurerstrasse 190, CH–8057 Zürich, Switzerland}
\affiliation{Paul Scherrer Institut, CH–5232 Villigen PSI, Switzerland}

\author{Andreas Crivellin\orcidB{}}
\email{andreas.crivellin@cern.ch}
\affiliation{Physik-Institut, Universität Zürich, Winterthurerstrasse 190, CH–8057 Zürich, Switzerland}
\affiliation{Paul Scherrer Institut, CH–5232 Villigen PSI, Switzerland}

\author{Srimoy Bhattacharya}
\email{bhattacharyasrimoy@gmail.com}
\affiliation{School of Physics and Institute for Collider Particle Physics, University of the Witwatersrand,
Johannesburg, Wits 2050, South Africa}

\author{Bruce Mellado}
\email{bmellado@mail.cern.ch}
\affiliation{School of Physics and Institute for Collider Particle Physics, University of the Witwatersrand,
Johannesburg, Wits 2050, South Africa}
\affiliation{iThemba LABS, National Research Foundation, PO Box 722, Somerset West 7129, South Africa}

\begin{abstract}
{In recent years, several hints for new scalar particles have been observed by the Large Hadron Collider at CERN. In this context}  we recast and combine the CMS and ATLAS analyses of the Standard Model Higgs boson decaying to a pair of $W$ bosons in order to search for low-mass resonances. We provide limits on the corresponding cross section assuming direct production via gluon fusion. For the whole range of masses we consider (90$\,$GeV to 200$\,$GeV), the observed limit on the cross section turns out to be weaker than the expected one. Furthermore, at $\approx95\,$GeV the limit is weakest and a new scalar decaying into a pair of $W$ bosons (which subsequently decay leptonically) with a cross section $\approx0.5\,$pb is preferred over the Standard Model hypothesis by $\gtrsim 2.5\,\sigma$. In light of the {previously existing excesses in other channels at similar masses, this strengthens the case for such a new Higgs boson and gives room} for the scalar candidate at 95$\,$GeV decaying into $W$ bosons.
\end{abstract}
\maketitle

\section{Introduction} 
The Standard Model (SM) of particle physics describes very successfully the fundamental constituents of matter as well as their interactions. It has been extensively tested and verified at both the precision and high-energy frontiers~\cite{ParticleDataGroup:2020ssz,HFLAV:2019otj,ALEPH:2005ab} with the discovery of the Brout-Englert-Higgs boson~\cite{Higgs:1964ia,Englert:1964et,Higgs:1964pj,Guralnik:1964eu} at the LHC~\cite{Aad:2012tfa,Chatrchyan:2012ufa} providing the final missing puzzle piece, as this 125 GeV boson ($h$) has the properties predicted by the SM to a good approximation. 

However, this does not exclude the existence of additional scalar bosons as long as their role in the breaking of the SM electroweak gauge symmetry is sufficiently small. In fact, searches for new resonances at the LHC (see, e.g,.Ref.~\cite{ATLAS:2022jsi} for a recent review), including additional scalar bosons~\cite{Naryshkin:2021ryt}, have been intensified since the Higgs boson discovery. While the LHC experiments ATLAS and CMS did not observe unequivocally the production of such a new particle, interesting hints for a new scalar with a mass around 95$\,$GeV~\cite{LEPWorkingGroupforHiggsbosonsearches:2003ing,CMS:2018cyk,CMS:2022rbd,CMS:2022tgk,Cao:2016uwt,Biekotter:2019kde,Crivellin:2017upt,Haisch:2017gql,Fox:2017uwr,Heinemeyer:2018jcd,Biekotter:2022jyr,Iguro:2022dok}, 151$\,$GeV~\cite{vonBuddenbrock:2017gvy,ATLAS:2021jbf,Crivellin:2021ubm,Richard:2021ovc,Fowlie:2021ldv}, and 680$\,$GeV~\cite{CMS:2017dib,ATLAS:2021uiz,CMS:2022tgk,Consoli:2021yjc} arose, as well as anomalies in multilepton final states~\cite{vonBuddenbrock:2017gvy,vonBuddenbrock:2019ajh,vonBuddenbrock:2020ter,Hernandez:2019geu,Fischer:2021sqw}.\footnote{This includes hints for the enhanced nonresonant production (i.e.~not originating from the direct two-body decay of a new particle) of different-flavor opposite-sign dileptons which can be explained by the decay of a neutral scalar with a mass between $130\,$GeV and $170\,$GeV~\cite{vonBuddenbrock:2017gvy} decaying into pairs of $W$ bosons~\cite{vonBuddenbrock:2016rmr,vonBuddenbrock:2018xar}. In fact, assuming the decay of a scalar into $WW$, Ref.~\cite{vonBuddenbrock:2017gvy} reported a combined best fit of $150\pm5$\, GeV. The range of interest is widened here in order to accommodate other decay mechanisms (such as associated production, i.e.~the production in association with an additional particle) and to cover the interesting range around the $95\,$GeV.}  

{Therefore, the question arises, if these hints for new scalars are accompanied and supported by signals of their decays into $W$ bosons.} While there is even an excess in $WW$ searches at a mass around $650\,$GeV in the vector-boson fusion category, and a weaker than expected limit around 150$\,$GeV in the gluon fusion category in the latest CMS analysis~\cite{CMS:2022bcb}, the mass range below 300$\,$GeV is not covered by the corresponding ATLAS analysis~\cite{ATLAS:2022qlc}. Furthermore, Ref.~\cite{CMS:2022bcb} stops at 115$\,$GeV and therefore does not cover the interesting region around 95$\,$GeV, while Ref.~\cite{CMS:2019bnu} searched down to masses of 100$\,$GeV, but this analysis was done with only 35.9$\,$fb$^{-1}$ of integrated luminosity. 

\begin{figure}[h!]
    \includegraphics[width=0.7\linewidth]{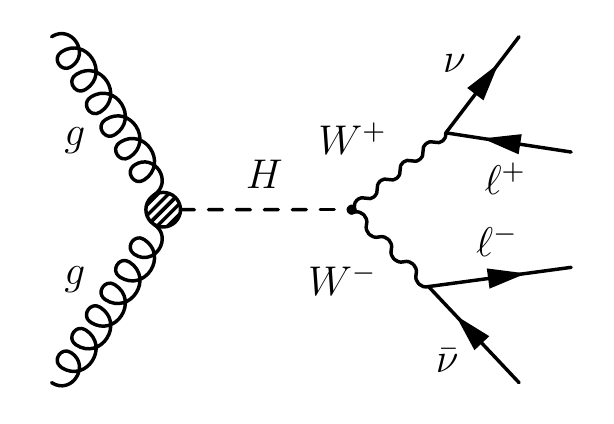}
    \caption{Feynman diagram showing the direct production via gluon fusion of a new scalar $H$ with mass $m_H$ decaying into a pair of $W$ bosons that then decay leptonically, resulting in the signatures studied in Refs.~\cite{CMS:2022uhn,ATLAS:2022ooq} in the context of the SM Higgs boson.}
    \label{fig:Feynman}
\end{figure}

Therefore, in this article, we will fill the gap by recasting the LHC analyses of ATLAS and CMS for the SM Higgs boson decaying into $WW$~\cite{CMS:2022uhn,ATLAS:2022ooq} and combining them in a global fit. This has the advantage that it is well suited for scalar masses in the range of 90$\,$GeV to 200$\,$GeV, and that for this mode, both ATLAS and CMS analyses with the full run-2 dataset, corresponding to 139$\,$fb$^{-1}$ and  138$\,$fb$^{-1}$ of integrated luminosity, respectively, are available. For this purpose, we will assume that the new neutral scalar $H$ is produced directly via gluon fusion and decays with a sizable branching fraction into $W$ pairs that subsequently decay leptonically (see Fig.~\ref{fig:Feynman}).

\section{Simulation and Validation}

We consider a new neutral scalar $H$ with mass $m_H$ at the LHC, that is produced directly via gluon fusion and decays dominantly into a pair of $W$ bosons (one of which can be off shell) which subsequently decay leptonically. Note that such a setup can be naturally obtained if the scalar is the neutral component of an $SU(2)_L$ triplet~\cite{Rizzo:1990uu,Chardonnet:1993wd} with hypercharge~0, which, at tree level, disregarding mixing with the SM Higgs boson, only decays into a pair of $W$ bosons. Interestingly, the vacuum expectation value of this field contributes positively to the $W$ mass at tree level~\cite{Blank:1997qa,FileviezPerez:2022lxp} and can thus provide a natural explanation~\cite{Rizzo:2022jti,Wang:2022dte,Cheng:2022hbo,Song:2022jns} of the CDF~II measurement~\cite{CDF:2022hxs}, which lies above the SM prediction~\cite{deBlas:2021wap,Bagnaschi:2022whn}. Note that the masses of the components of the real $SU(2)_L$ triplet scalar field are largely unconstrained by LHC searches~\cite{Chabab:2018ert,Cacciapaglia:2022bax}.

\begin{figure}
    \includegraphics[width=1\linewidth]{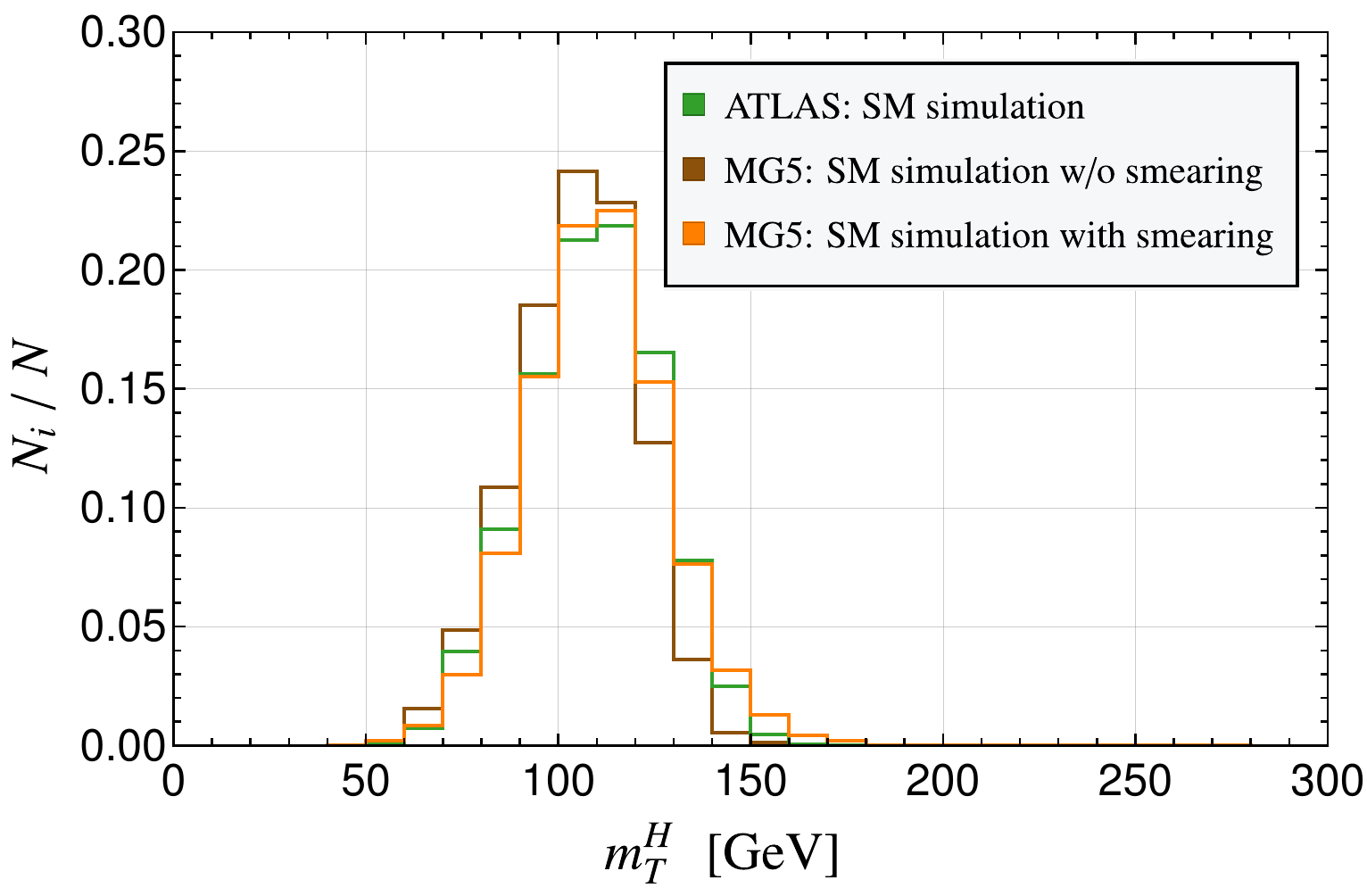}
    \caption{SM Higgs boson signal in the ATLAS analysis and our simulation with and without smearing (and shift) normalized to the total number of events of the respected simulation. One can see that once smearing and a shift are included, the agreement is very good between the two distributions. While the corresponding CMS plot is not shown, the results are very similar.}
    \label{fig:Smearing}
\end{figure}

We simulated in this setup the process $pp\to H\to W W^{(*)}\to \ell^+\ell^-\nu\bar{\nu}$, with $\ell=e,\mu,\tau$, and the tau lepton subsequently decaying, using {\tt MadGraph5aMC@NLO} (MG5)~\cite{Alwall:2014hca}, {\tt Pythia8.3}~\cite{Sjostrand:2014zea}, and {\tt Delphes}~\cite{deFavereau:2013fsa}. For each point in parameter space that we will consider in the following, we generated a sample containing one million events. In order to validate and correct our fast simulation, we first simulated the SM Higgs boson signal, i.e.~$gg\to h\to W W^{(*)}\to \ell^+\ell^-\nu\bar{\nu}$, and compared the result to the ATLAS one for the SM Higgs boson signal given as a function of the transverse mass $m_T$ 
in Fig.~11 in Ref.~\cite{ATLAS:2022ooq}.\footnote{Note that the $m_T$ definitions of ATLAS and CMS are slightly different (for instance, they contain details on the missing energy) and that we do not use the di-lepton invariant mass here as it is fully correlated to the transverse mass, the latter, however, containing more information.} For this, we normalized the events per bin $N_i$ to the total number of events $N$ and then calculated the sum of the square of the differences between the two simulations of all bins~$\Delta=\sum_i (N_i^{\rm ATLAS}-N_i^{\rm MG5})^2$, where MG5 stands for our {\tt MadGraph5aMC@NLO} simulation. It turns out that to better match the $m_T$  distribution of ATLAS with our fast simulation, a smearing on the missing transverse energy $E_T$ to broaden the $m_T$ spectrum is necessary, as well as a shift of the $m_T$ to adjust the position of the peak. We thus uniformly generate random numbers $r$ and $\phi$ on a disk in the $x-y$ plane, i.e., ~$k \times r (\cos \phi+\sin \phi)$ with $k$ in units of GeV, $r$ between 0 and 1 and $\phi$ between 0 and $2\pi$. We then add the resulting values to the missing $E_T$ generated. In fact, we found that the best fit, i.e.~minimal value for $\Delta$ is obtained for $k\approx\,$20$\,$GeV. In addition, a shift of $\approx\,$3.5$\,$GeV on the transverse mass leads to a very good agreement between our simulation and the one of ATLAS (see Fig.~\ref{fig:Smearing}).

Next, we look at the production cross section and the efficiency of our simulation compared to the ATLAS one. First, note that in the ATLAS graph, the fitted signal (i.e.~the one that agrees best with data, not taking into account the overall normalization from the SM theory prediction) is shown, such that one has to rescale the number of events by dividing by 1.21. We then corrected for the leading order MG5 simulation using an effective $ggh$ coupling. The resulting production cross section is 17.62$\,$pb, while including NNLO corrections,\footnote{Since we consider the 0-jet category, hard jet emission is vetoed. Therefore, $\alpha_s$ corrections are only relevant for the production cross section, which is however fitted in our approach.} the CERN yellow report~\cite{LHCHiggsCrossSectionWorkingGroup:2016ypw} quotes 48.57$\,$pb. We also corrected for the simulation efficiency, i.e., ~the percentage of events left after applying the cuts, which in our analysis is $\epsilon_{\rm MG5}\approx0.017$ while ATLAS finds $\epsilon_{\rm ATLAS}=0.011$, being in reasonable agreement. 

\begin{figure}[t]
\centering
    \includegraphics[width=0.98\linewidth]{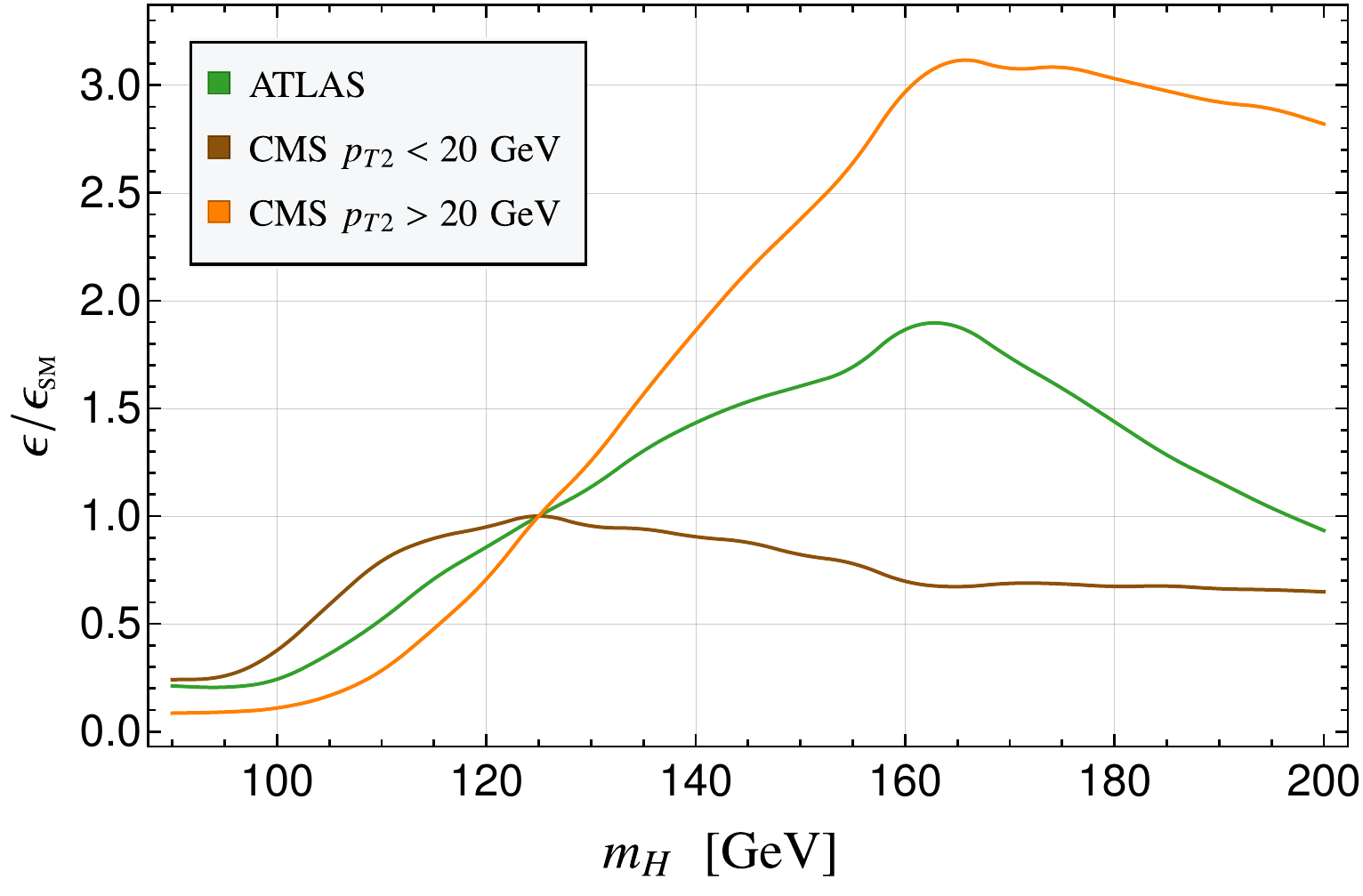}
    \caption{Dependence of the efficiency of our simulation, normalized to the one at 125 GeV, as a function of the mass of the new scalar for the ATLAS analysis and the two CMS categories.}
    \label{fig:efficiency}
\end{figure}

We proceeded in a similar way for the CMS analysis, both for the $p_{T2}<20\,$GeV and the $p_{T2}>20\,$GeV categories (where $p_{T2}$ stands for the transverse momentum of the subleading lepton) shown in Fig.~1 of Ref.~\cite{CMS:2022uhn}. Here, a smearing of $k=30\,$GeV gives the best fit, while a shift is not necessary. For the production cross section, the same correction factor applies, while for the combined efficiency ($p_{T2}<20\,$GeV and $p_{T2}>20\,$GeV category), we find $\epsilon_{\rm MG5}\approx0.019$ while CMS finds $\epsilon_{\rm CMS}\approx0.012$, again in reasonable agreement.\footnote{{The difference in the efficiency of our fast simulation compared to the full simulation of ATLAS and CMS can be explained by pileup reducing the (unrealistically) high electron and muon efficiency of $95\%$ in {\tt Delphes}, compared to the one of ATLAS and CMS for medium energetic leptons~\cite{Jain:2021bis} as well as the full jet veto used.}} 
\newline The dependence of the efficiency (relative to the SM) on $m_H$ is shown in Fig.~\ref{fig:efficiency}.
For the BSM analysis, we will then apply the correction factors, as well as the smearing, determined from the SM Higgs boson. For the shift in $m_T$, we assumed that it is proportional to the scalar mass $m_H$. 

\section{Analysis}

For the ATLAS analysis~\cite{ATLAS:2022ooq}, we digitized the data points as well as the backgrounds and the SM Higgs boson signal for the 0-jet category\footnote{Here, we do not include the 1-jet and 2-jet categories. This is motivated by the fact that the multilepton anomalies include the production of opposite-sign leptons in association with $b$jets, thus contaminating the control samples used to normalize the $t\overline{t}$ backgrounds in the 1-jet and 2-jet categories~\cite{vonBuddenbrock:2017gvy,vonBuddenbrock:2019ajh}. Note that these categories are anyway less sensitive than the 0-jet one for the gluon-gluon fusion signal considered here.} as a function of the transverse mass (Fig.~11 in the ATLAS paper). Concerning the latter, ATLAS scaled the theory prediction by 1.21 in order to obtain the best fit. As we study BSM effects, we, therefore, divided this contribution by this factor. For the statistical errors, we used the square root of the measured number of events per bin. Concerning the systematic error, one can see that there is a strong anti-correlations among the different background signals (including the SM Higgs boson signal) in Table~5 of the ATLAS paper. As the details of the (anti-)correlations are not given in the ATLAS paper, and the error on the Mis-Id background matches the total error, we chose this to be the experimental systematic error, also because it is reasonably the least correlated one with respect to the other backgrounds (which depend mostly on the detector efficiencies for leptons). Concerning the theory uncertainty, we included a 7\% error on the SM Higgs boson signal {(see Table~6 in Ref.~\cite{CMS:2022uhn})}. Furthermore, we assumed both systematic errors to be uncorrelated from each other but fully correlated among the different bins. 

\begin{figure}
    \centering
    \includegraphics[width=1\linewidth]{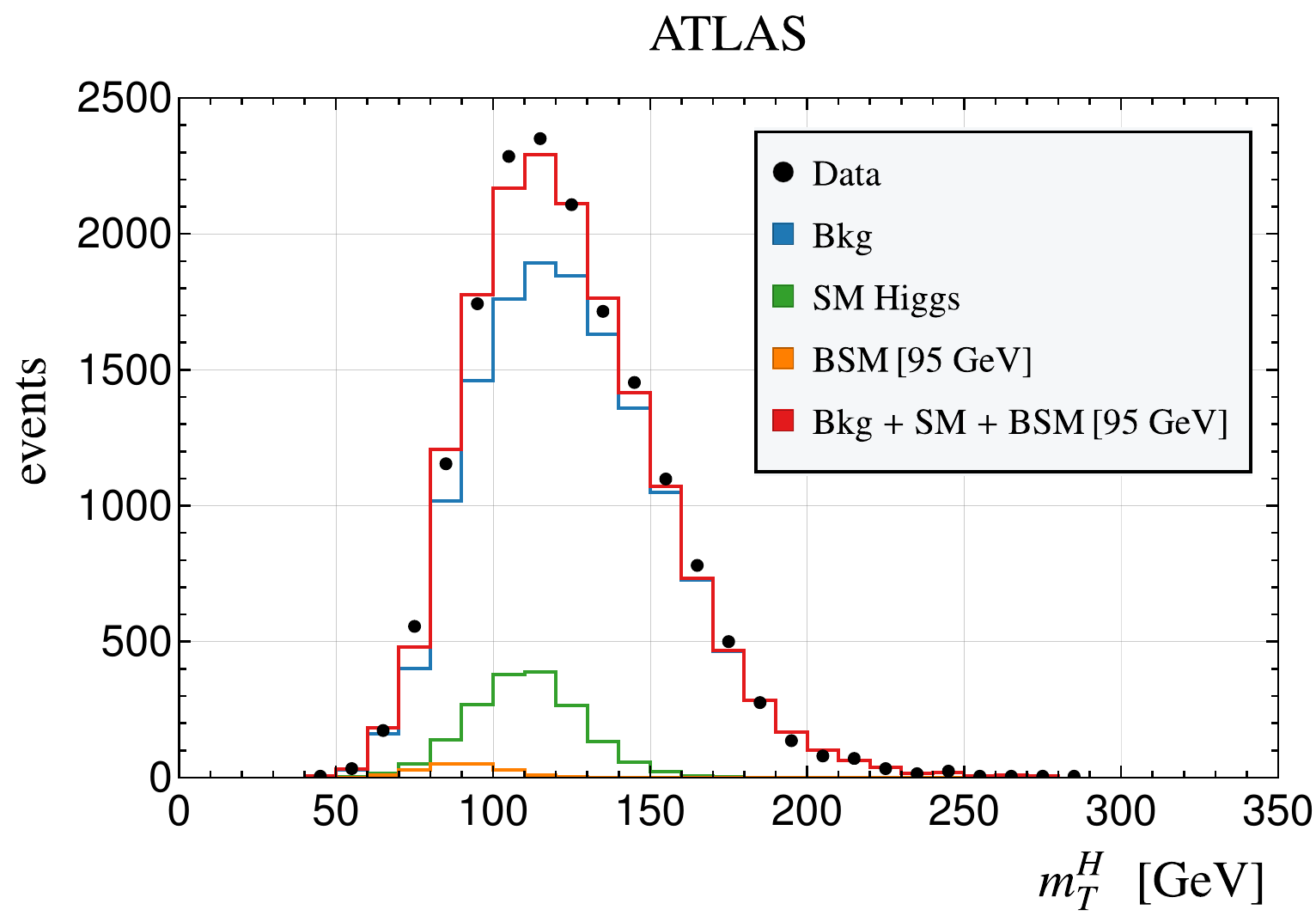}
    \includegraphics[width=1\linewidth]{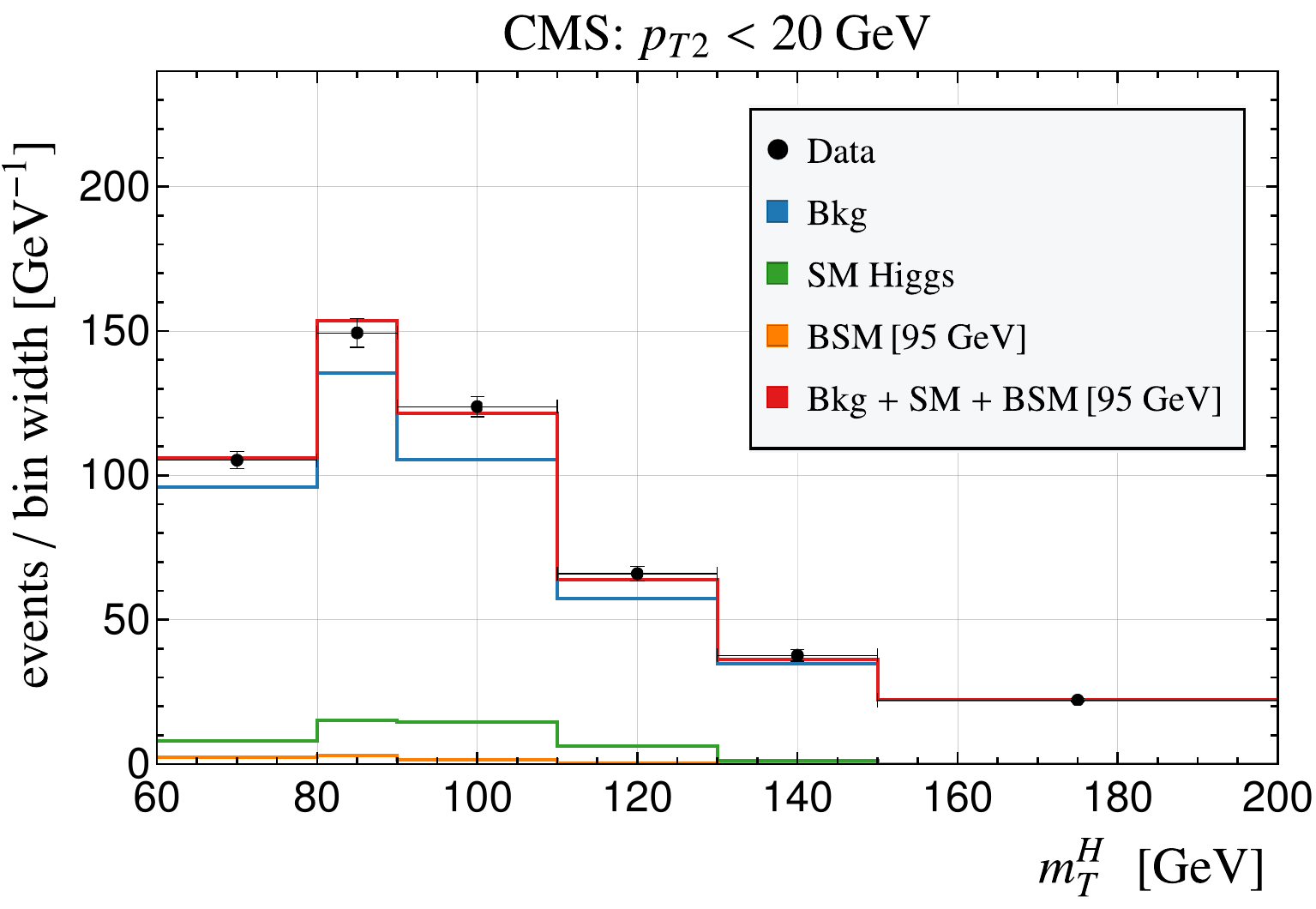}
    \caption{Results of the fit to the ATLAS and CMS analyses of $pp\to H\to WW^{(*)}\to \ell^+\ell^-\nu\bar{\nu}$ for the case of a new scalar with a mass of $95\,$GeV. Only the 0-jet category is used here (see text) and the CMS category with $p_T>20\,$GeV is not shown, due to the very small efficiency.}
    \label{fig:h95compATLAS}
\end{figure}

Analogously to the ATLAS procedure, we digitized the $m_T$ distributions for the $p_{T2}<20\,$GeV and $p_{T2}>20\,$GeV categories in Fig.~1 of Ref.~\cite{CMS:2022uhn}. However, CMS uses a different method for determining background and signal, namely a combined fit to data. Therefore, in the presence of a BSM signal, we allowed for refitting the SM background (including the SM Higgs boson signal) by a common factor $\mu_{\scriptscriptstyle  \rm BKG}$, which can however be different for the two categories $p_{T2}<20\,$GeV and the $p_{T2}>20\,$GeV. This at the same time takes into account the experimental systematic uncertainties of the main $WW$ background and the SM Higgs boson. Since for CMS, the systematic error on the nonprompt background is not given, we used $13\%$ as for the ATLAS analysis. On top of this, we included $7\%$ systematic theory error of the SM Higgs boson signal, the latter fully correlated among the $p_{T2}<20\,$GeV and the $p_{T2}>20\,$GeV categories and with the theory uncertainty for the ATLAS analysis.

\begin{figure*}[t]
\centering
    \includegraphics[width=0.49\linewidth]{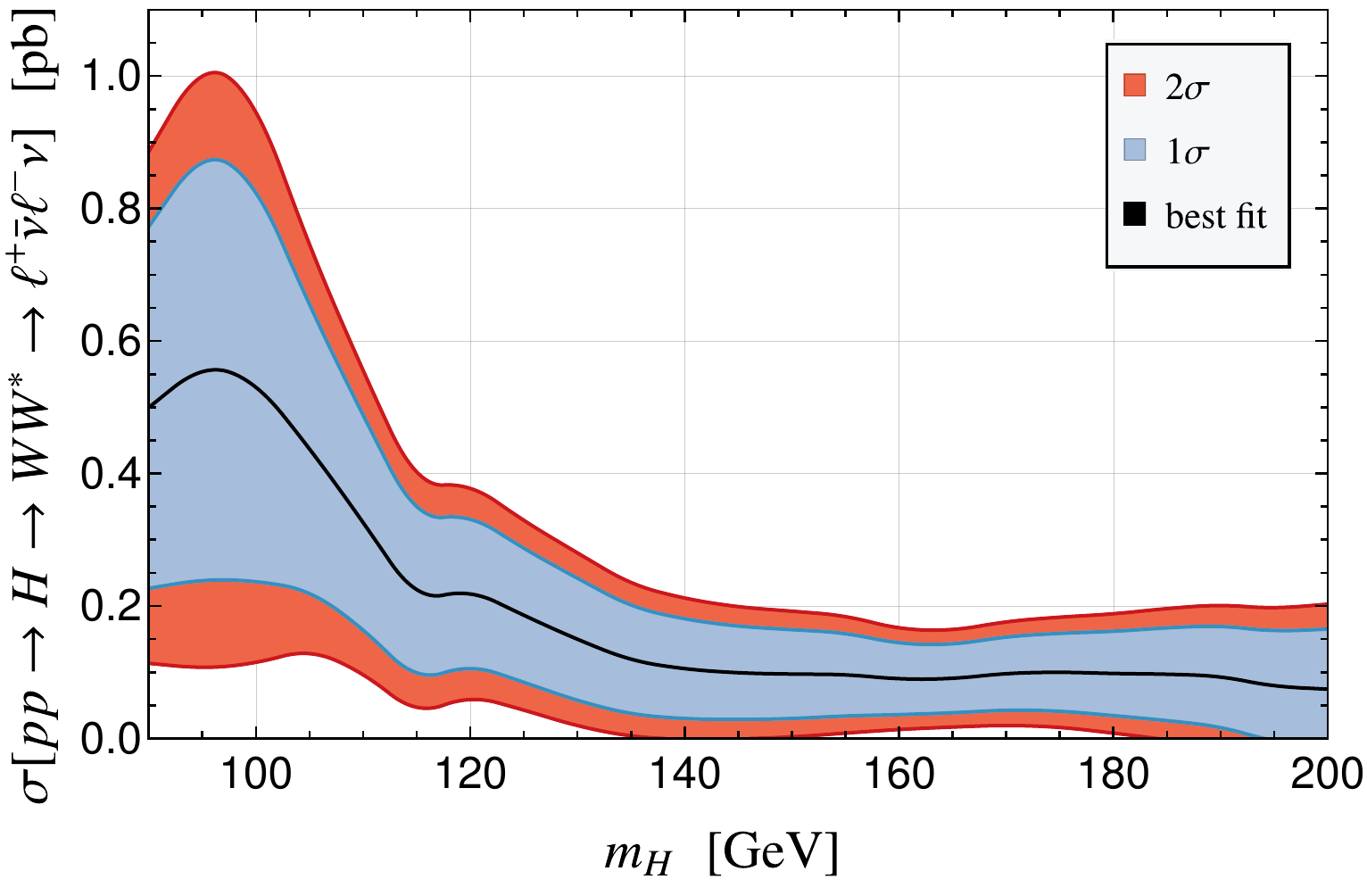}
    \includegraphics[width=0.49\linewidth]{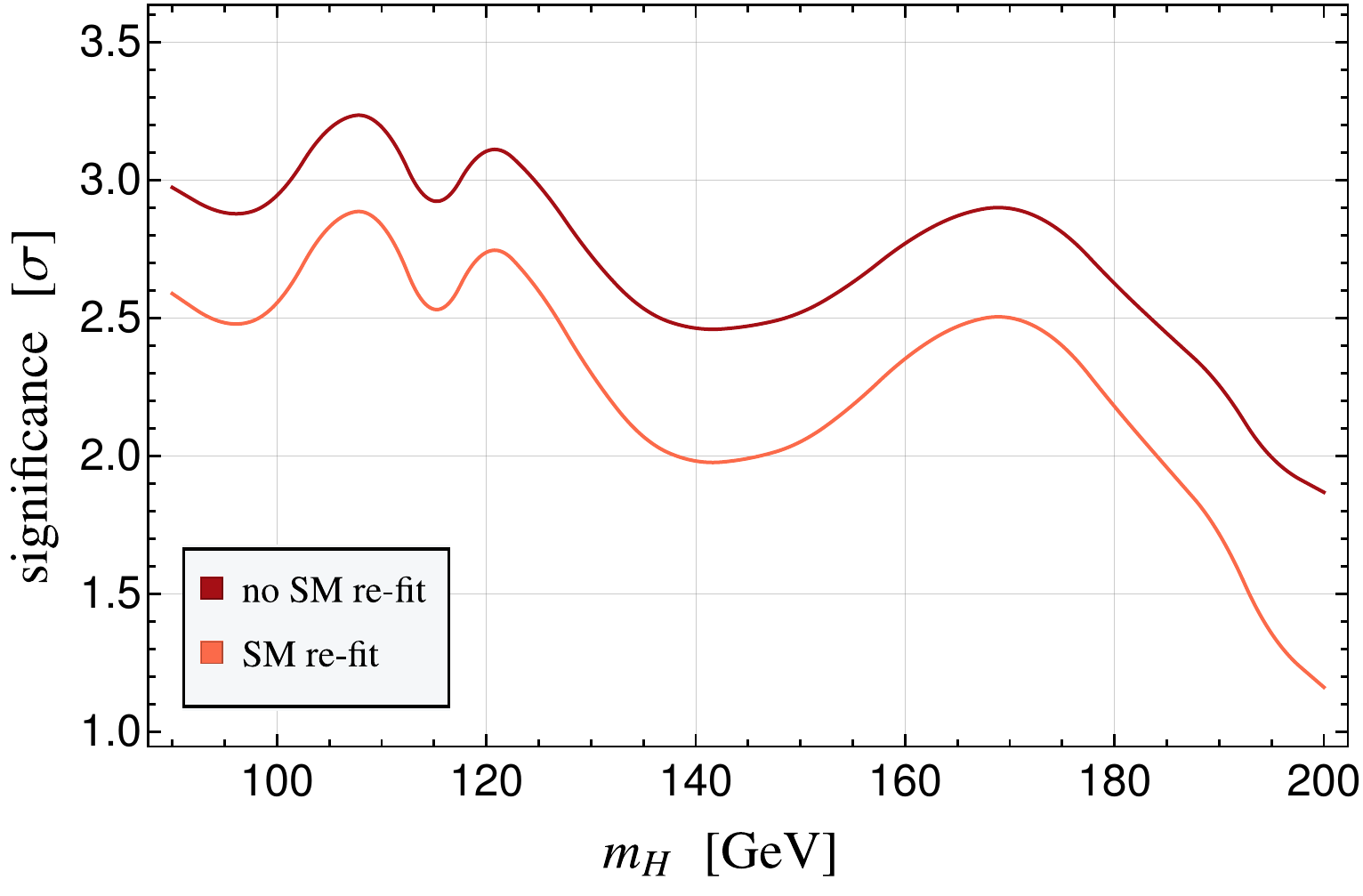}
\caption{Left: Preferred range for $\sigma[pp \rightarrow H \rightarrow WW^{(*)} \rightarrow \ell^+ \bar{\nu}\ell^- \nu]$ from the combined fit to ATLAS and CMS data as a function of $m_H$, covering the range from $90\,$GeV up to $200\,$GeV. The largest cross section is allowed at around $95\,$GeV. Right: Significance for a BSM signal using the two different methods for treating the SM background of CMS analyses. }
    \label{mhscan}
\end{figure*}

The statistical model for the combined analysis is then built up with binned templates from observed data and expectations, including a possible BSM signal. In order to obtain the best-fit value of BSM signal strength, a simultaneous fit based on $\chi^2$ distribution is performed. For this, we calculate  a common $\chi^2$ depending on the BSM signal,
\begin{equation}
    \chi^2_{\rm BSM} = [N_i^{\rm data} - N_i^{\rm theory} ] \; \Sigma_{ij}^{-1} \; [N_j^{\rm data} - N_j^{\rm theory} ]\,,
\end{equation}
where $\Sigma_{ij}$ is the covariance matrix, $N_i^{\rm data}$ is the number of measured events per bin, and
\begin{equation}
    N_i^{\rm theory} = \rho_{\scriptscriptstyle \rm BKG} (N_i^{\scriptscriptstyle \rm SM} + N_i^{\scriptscriptstyle \rm BKG}) + \mu_{\scriptscriptstyle \rm BSM}N_i^{\scriptscriptstyle \rm BSM},
\end{equation}
is composed of the background (BKG) events, the number of events expected within the SM, and the BSM component, with each contribution weighted by a respective fit parameter. We normalized the number of BSM events to the number of events within the SM, i.e., $\sum_i N_i^{\rm SM}=\sum_i N_i^{\rm BSM}$ such that
\begin{equation}
 \mu_{\scriptscriptstyle \rm BSM}=\frac{\sigma[pp \rightarrow H \rightarrow WW^{(*)} \rightarrow \ell^+ \bar{\nu}\ell^- \nu]}{\sigma[pp \rightarrow h \rightarrow WW^* \rightarrow \ell^+ \bar{\nu}\ell^- \nu]}\,.
\end{equation}
While in Table~\ref{tab:resmass150}, we will give $\mu_{\scriptscriptstyle \rm BSM}$ also separately for ATLAS, CMS with $p_{T2}<20\,$GeV and CMS with $p_{T2}>20\,$GeV, in our final combined fit, we will require it to be equal for all three categories.

\section{Results}

By minimizing the global $\chi^2_{\rm BSM}$ function, a best-fit value of $\mu_{\scriptscriptstyle \rm BSM}$ can be derived, and the corresponding $\chi^2$ value is then compared to the SM value $\chi_{\rm SM}^2$. For the latter, a subtlety arises in the case of the CMS analyses: one can either use the value obtained directly from the CMS plots or allow for refitting the backgrounds, as done for the BSM analysis. While the latter option is more conservative, the first option seems more appropriate in the case of a nonzero BSM signal. We will therefore give both numbers in the following.

First, let us look at the results for the particularly interesting case of $m_H=95\,$GeV and $m_H=150\,$GeV, which are motivated by the anomalies mentioned in the introduction. The result is illustrated in Fig.~\ref{fig:h95compATLAS} for a mass of $m_H=95\,$GeV, and the numbers for both cases are given in Table~\ref{tab:resmass150}, where both the individual as well as the combined fit results are shown. In the leftmost part of the table, one can find the best-fit values for the parameters. The middle (rightmost) parts correspond to results in which the $\chi^2$ for the SM hypothesis is obtained with (without) refitting the background and the SM signal for the CMS analyses. 

Finally, we show the preferred range of the cross section of $pp\to H\to WW^{(*)} \to\ell^+\nu\ell^-\bar\nu$ as a function of $m_H$ from $90\,$GeV up to $200\,$GeV in Fig.~\ref{mhscan}, where we scanned over the mass in steps of $5\,$GeV and then interpolated. The black line denotes the best fit while blue and red correspond to the $1\sigma$ and $2\sigma$ regions, respectively. The largest possible cross section is allowed for $\approx 95\,$GeV and also at larger masses there is room for a BSM signal. Note that in the left plot of Fig.~\ref{mhscan}, we defined the $1\sigma$ and $2\sigma$ regions w.r.t.~the best-fit values of the BSM scenario, allowing for a refit of the SM background for CMS even in case of a vanishing signal. Therefore, these regions correspond to the conservative approach discussed above.

\section{Conclusions and Outlook}

\SetTblrInner{rowsep=0.15cm}
\begin{table*}[!t]
\begin{center}
\begin{tblr}{colspec={|[1.5pt]Q[c,2.8cm,lg]|[1.5pt]Q[c,1.2cm]|Q[c,1.2cm]|Q[c,1.2cm]|Q[c,1.2cm]||Q[c,1.2cm]|Q[c,1.2cm]||Q[c,1.2cm]|Q[c,1.2cm]|[1.5pt]},row{1,6} = {bg = lg}, cell{1,6}{1} = {bg = white}}
\hline[1.5pt]
 $\textit{m}_{\textit{H}} = 95\,$GeV & $\rho_{\scriptscriptstyle \text{BKG}}^{p_{T2}< 20}$ & $\rho_{\scriptscriptstyle \text{BKG}}^{p_{T2} > 20}$ & $\mu_{\scriptscriptstyle \rm BSM}$   & $\chi^2_{\scriptscriptstyle \text{BSM}}$ & $\chi_{\scriptscriptstyle \text{SM}}^{2,\text{refit}}$ & Sig.$ ^{\text{refit}}$ &  $\chi^2_{\scriptscriptstyle \text{SM}}$ & Sig. \\
 \hline[1pt]
ATLAS                           &       &       & 0.7   & 49.0  & 57.7   & 3.0$\,\sigma$ & 57.7 & 3.0$\,\sigma$     \\
\hline
 CMS $ p_{T2} < 20\,$GeV        & 1.01  &       & 0.0   &  5.5  &  5.5   & 0.0$\,\sigma$  & 6.8  & 1.2$\,\sigma$     \\
 \hline
 CMS $ p_{T2} > 20\,$GeV        &       & 1.01  & -3.5   &  6.2  & 9.0    & -  & 9.1  & -      \\
 \hline
 Combined fit                   & 1.00  & 1.00  & 0.5    & 65.4  & 72.2   & 2.6$\,\sigma$ & 73.3 & 2.8$\,\sigma$       \\ 
 \hline[1pt]
 \hline[0pt]
 \hline[1pt]
$\textit{m}_{\textit{H}} = 150\,$GeV & $\rho_{\scriptscriptstyle \text{BKG}}^{p_{T2}< 20}$ & $\rho_{\scriptscriptstyle \text{BKG}}^{p_{T2} > 20}$ & $\mu_{\scriptscriptstyle \rm BSM}$   & $\chi^2_{\scriptscriptstyle \text{BSM}}$ & $\chi_{\scriptscriptstyle \text{SM}}^{2,\text{refit}}$ & Sig.$^{\text{refit}}$ &  $\chi^2_{\scriptscriptstyle \text{SM}}$ & Sig. \\
\hline[1.5pt]
ATLAS                           &       &       & 0.1   & 54.5  & 57.7   & 1.8$\,\sigma$ & 57.7 & 1.8$\,\sigma$     \\
\hline
 CMS $ p_{T2} < 20\,$GeV        & 0.97  &       & 0.6   &  1.5  &  5.5   & 2.0$\,\sigma$ & 6.8  & 2.3$\,\sigma$     \\
 \hline
 CMS $ p_{T2} > 20\,$GeV        &       & 0.99  & 0.2   &  8.0  & 9.0    & 1.0$\,\sigma$ & 9.1  & 1.0$\,\sigma$      \\
 \hline
 Combined fit                   & 1.01  & 0.99  & 0.1   & 67.2  & 72.2   & 2.2$\,\sigma$ & 73.3 & 2.5$\,\sigma$       \\ 
 \hline[1.5pt]
\end{tblr}
\caption{Fit results for the two cases $m_H = 95\,$GeV and $m_H = 150\,$GeV, motivated by the existing hints for new scalars at the LHC. {The significance is reported both with (Sig.$^{\text{refit}}$) and without (Sig.) refitting the background and the SM signal for the CMS analyses. The same notation is employed for $\chi_{\scriptscriptstyle \text{SM}}^{2,\text{refit}}$ and $\chi_{\scriptscriptstyle \text{SM}}^{2}$.} Note that the sizable value of $\mu_{\scriptscriptstyle \rm BSM}$ in the CMS $ p_T> 20\,$GeV category for the 95$\,$GeV case is due to the very small efficiency.}
\label{tab:resmass150}
\end{center}
\end{table*}
Motivated by the existing hints for new scalar particles with masses around $\approx 95\,$GeV and $\approx 151\,$GeV, we recast and combine the CMS and ATLAS analyses of the SM Higgs boson decaying into $W$ boson pairs to constrain light new scalars with a mass between 90$\,$GeV and 200$\,$GeV. In Fig.~\ref{mhscan} we show the preferred $1\sigma$ and $2\sigma$ ranges for the corresponding cross section. Note that for the whole range, the observed limit is weaker than the expected one, resulting in a preference for nonzero BSM contribution. While the allowed cross section is largest around $95$\, GeV, the global significance is only below $\approx\!2\,\sigma$. However, taking into account the existing hints for a $95$\,GeV scalar in $\gamma\gamma$, the look-elsewhere effect is removed, and the global significance of our $WW$ signal coincides with the local one of $\gtrsim 2.5\,\sigma$. Note that while for a $151\,$GeV scalar there is already room for a positive signal in our setup with direct production, its production in association with missing energy is suggested by Refs.~\cite{ATLAS:2021jbf,ATLAS:2023omk} ($H\rightarrow \gamma\gamma$) and Refs.~\cite{ATLAS:2020fgc,ATLAS:2022bzt} ($H\rightarrow WW \rightarrow 4q, \ell\nu 2q$). While such an associate production will broaden the values for $m_T$, further increasing the significance, the quantification of this effect is outside the scope of this paper.

Due to the absence of a $ZZ\to 4\ell$ signal in the LHC analyses, our results suggest that the new scalar could be the neutral component of an $SU(2)_L$ triplet with hypercharge 0, that, at tree level and in the absence of mixing, only decays to a pair of $W$ bosons. This observation is interesting in light of the fact that this field can at the same time naturally account for the $W$ mass measurement of the CDF~II Collaboration, in case its vacuum expectation value is around a few GeV.

\begin{acknowledgments}
We thank Mukesh Kumar, Shuiting Xin, Salah-Eddine Dahbi and Saiyad Ashanujjaman for useful discussions. The work of A.C.~is supported by a professorship grant of the Swiss National Science Foundation (Grant No.\ PP00P21\_76884). B.M.~gratefully acknowledges the South African Department of Science and Innovation through the SA-CERN program, the National Research Foundation, and the Research Office of the University of the Witwatersrand for various forms of support.
\end{acknowledgments}

\bibliography{apssamp}

\end{document}